\documentclass[aps,showpacs,superscriptaddress,reprint,prb]{revtex4-1}

\usepackage{amsthm}
\usepackage{physics}
\usepackage{amsfonts}
\usepackage{amsmath}
\usepackage{siunitx}
\usepackage{multirow}
\usepackage{graphicx}
\usepackage{caption}
\usepackage[usenames,dvipsnames,svgnames,table]{xcolor}
\usepackage{dcolumn}
\usepackage{bm}
\usepackage{gensymb}
\newcommand{\cc}[1]{\textcolor{black}{#1}}

\begin{document}

\title{High resolution resonant inelastic EUV scattering from orbital excitations in a Heisenberg antiferromagnet}
\author{Antonio Caretta}
\affiliation{Elettra-Sincrotrone Trieste S.C.p.A. Strada Statale 14 - km 163.5 in AREA Science Park 34149 Basovizza, Trieste, Italy}
\author{Martina Dell'Angela}
\affiliation{Elettra-Sincrotrone Trieste S.C.p.A. Strada Statale 14 - km 163.5 in AREA Science Park 34149 Basovizza, Trieste, Italy}
\affiliation{CNR-IOM, Strada Statale 14 - km 163.5 in AREA Science Park 34149 Basovizza, Trieste, Italy}
\author{Yi-De Chuang}
\affiliation{Advanced Light Source, Lawrence Berkeley National Laboratory, Berkeley, California 94720, USA}
\author{Alexandra M. Kalashnikova}
\affiliation{Ioffe Physical-Technical Institute, Politekhnicheskaya 26, 194021 St. Petersburg, Russia}
\author{Roman V. Pisarev}
\affiliation{Ioffe Physical-Technical Institute, Politekhnicheskaya 26, 194021 St. Petersburg, Russia}
\author{Davide Bossini}
\affiliation{Institute for Photon Science and Technology, Graduate School of Science, The University of Tokyo 7-3-1 Hongo, Bunkyo-ku, Tokyo 113-0033, Japan}
\author{Florian Hieke}
\affiliation{Physics Department and Center for Free-Electron Laser Science, Hamburg University, 22607 Hamburg, Germany}
\author{Wilfried Wurth}
\affiliation{Physics Department and Center for Free-Electron Laser Science, Hamburg University, 22607 Hamburg, Germany}
\affiliation{Photon Science, DESY, Notkestr. 85, 22607 Hamburg, Germany}
\author{Barbara Casarin}
\affiliation{Department of Physics, University of Trieste, Via A. Valerio 2, 34127 Trieste, Italy}
\author{Roberta Ciprian}
\affiliation{Elettra-Sincrotrone Trieste S.C.p.A. Strada Statale 14 - km 163.5 in AREA Science Park 34149 Basovizza, Trieste, Italy}
\author{Fulvio Parmigiani}
\affiliation{Department of Physics, University of Trieste, Via A. Valerio 2, 34127 Trieste, Italy}
\affiliation{Elettra-Sincrotrone Trieste S.C.p.A. Strada Statale 14 - km 163.5 in AREA Science Park 34149 Basovizza, Trieste, Italy}
\affiliation{International Faculty, University of Cologne, 50937 Cologne, Germany}
\author{Surge Wexler}
\affiliation{Department of Physics, New York University 4 Washington Place, New York, NY 10003}
\author{L. Andrew Wray}
\affiliation{Department of Physics, New York University 4 Washington Place, New York, NY 10003}
\author{Marco Malvestuto}
\email{marco.malvestuto@elettra.eu}
\affiliation{Elettra-Sincrotrone Trieste S.C.p.A. Strada Statale 14 - km 163.5 in AREA Science Park 34149 Basovizza, Trieste, Italy}

\begin{abstract}
We report the high resolution resonant inelastic EUV scattering study of quantum Heisenberg antiferromagnet KCoF$_3$. By tuning the EUV photon energy to cobalt M$_{23}$ edge, a complete set of low energy 3d spin-orbital excitations is revealed. These low-lying electronic excitations are modeled using an extended multiplet-based mean field calculation to identify the roles of lattice and magnetic degrees of freedom in modifying the RIXS spectral lineshape. We have demonstrated that the temperature dependence of RIXS features upon the antiferromagnetic ordering transition enables us to probe the energetics of short-range spin correlations in this material.   
\end{abstract}

\maketitle

\section{Introduction}
Resonant inelastic X-ray scattering (RIXS) spectroscopy\cite{Ament:2011jy} has emerged as powerful technique for studying the low-energy excitations in quantum materials\cite{Schlappa:2012hj,Monney:2013dfa,Dean:2012bd,Zhou:2013fu,Bisogni:2016gi,Ghiringhelli:2005kp}. \cc{In particular, it has been established that RIXS can probe the mutual coupling between distinct degrees of freedom and the corresponding collective excitations, such as electron-phonon\cite{Lee:2013ih} and spin-orbital\cite{Schlappa:2012hj} couplings.}
 
\cc{Nowdays, high energy resolutions ($\leq$25 meV) can be achieved by RIXS spectrometers in the soft and hard X-ray regime\cite{Anonymous:Od5c3xSg,Anonymous:gVo12F4s,Shvydko:2013cj,P01Desy:o-CF-nyY}.}  However, extending the RIXS spectroscopy into extreme ultraviolet (EUV)\cite{Wray:2015htb} regime can provide a wealth of benefits with respect to the soft X-ray energy range; for example, achieving superior energy resolution at moderate instrumentation resolving power and potentially offering simpler interpretations for the spectral features\cite{Augustin:2016gi,Wray:2015htb,Lee:2014ib,Wray:2015cu}. EUV RIXS at transition metal M edges is not only of basic relevance to condensed matter physics, but also presents a unique opportunity to investigate the elementary excitations in the time-domain\cite{Wray:2015htb}. In fact, recent RIXS experiments\cite{DellAngela:2016ua} had demonstrated that it was possible to record RIXS spectra using seeded EUV free electron laser source. With such, sub-picosecond time resolved RIXS experiments are now achievable thanks to the high FEL flux with unique time resolution. 

\begin{figure}[t!]
\captionsetup{justification=centerlast}
\centering
\includegraphics[width=0.8\columnwidth]{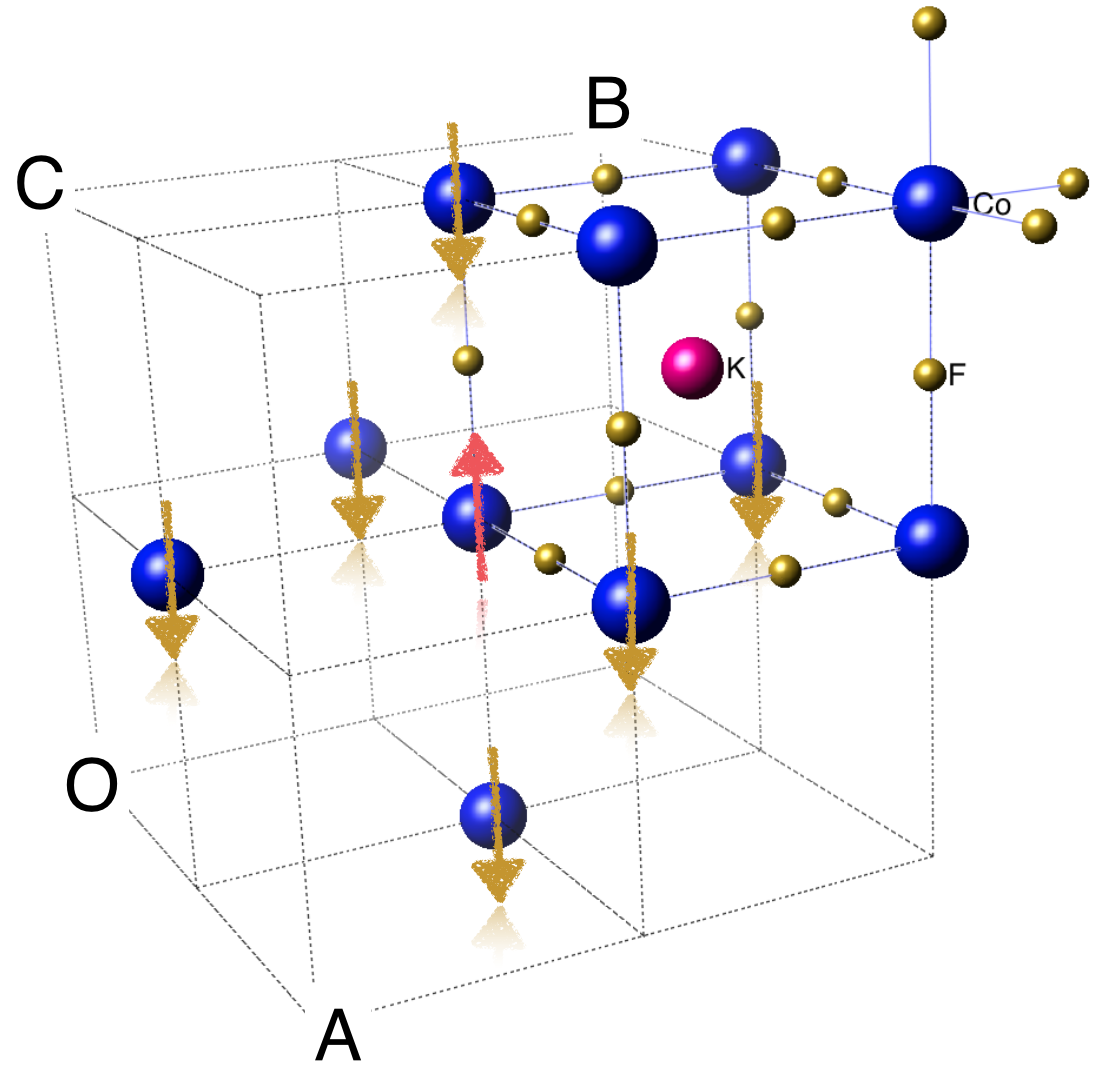}
\caption{The KCoF$_3$ lattice structure and a pictorial representation of the mean-field magnetic Hamiltonian. A single Co atom (red arrow) is coupled to 6 neighbours, each of which has the mean field spin moment (yellow arrow) for the second sublattice.}
\label{fig1}
\end{figure}

Here, we perform the high resolution EUV-RIXS (referred to as RIXS hereafter) measurements on antiferromagnet KCoF$_3$ to investigate how the magnetic ordering affects the low energy orbital excitations. KCoF$_3$ belongs to the actively researched family of insulating alkali metal fluorides KMF$_3$ (M = Cu, Mn, Fe, Co, Ni, and Zn)\cite{Maslen:1993hf,Okazaki:1961ds,Bondino:2009vf}, a cubic perovskite family that displays a variety of intriguing phenomena like high-temperature super-ionic behavior and interesting physical properties such as piezoelectric characteristics, ferromagnetism and non-magnetic insulator behaviors. They are the ideal candidates for studying the effect of super-exchange interaction on the structural, electronic and magnetic properties of a large class of ionic compounds. In addition, the highly ionic character, high crystallographic symmetry, and the low anion coordination of KMF$_3$ are expected to offer simpler modeling to describe their electronic and magnetic properties\cite{DeJongh:1974gb,DeJongh:1975ib,Dovesi:1997gp}. 
KCoF$_3$ has a G-type anisotropic antiferromagnetic structure with spins pointing along the c-axis\cite{Hirakawa:1964hi,Sharma:1976kt}. The magnetic unit cell contains two types of antiferromagnetically coupled neighbors and belongs to a non-symmorphic D$_h^4$ magnetic space group (see Fig. \ref{fig1}) \cite{Moch:1973jd,Okazaki:1961ds}. Upon establishing the antiferromagnetic ordering at T$_N$ = 115.3 K, the Jahn-Teller effect changes the crystal structure from cubic to tetragonal\citep{Dovesi:1997gp,Oleaga:2015fka}. 
We have studied the spectral line shape of orbital (dd) excitations across the N\'{e}el transition temperature. Using the extended atomic multiplet-based calculation to simulate the experimental RIXS spectra, we have determined the key energy scales for Co 3d orbitals. Our RIXS studies show that the onset of magnetic ordering, which quenches the spin-flip excitation channel, affects the overall energy of dd excitations. 

\begin{figure}[t!]
\captionsetup{justification=centerlast}
\centering
\includegraphics[width=1.00\columnwidth]{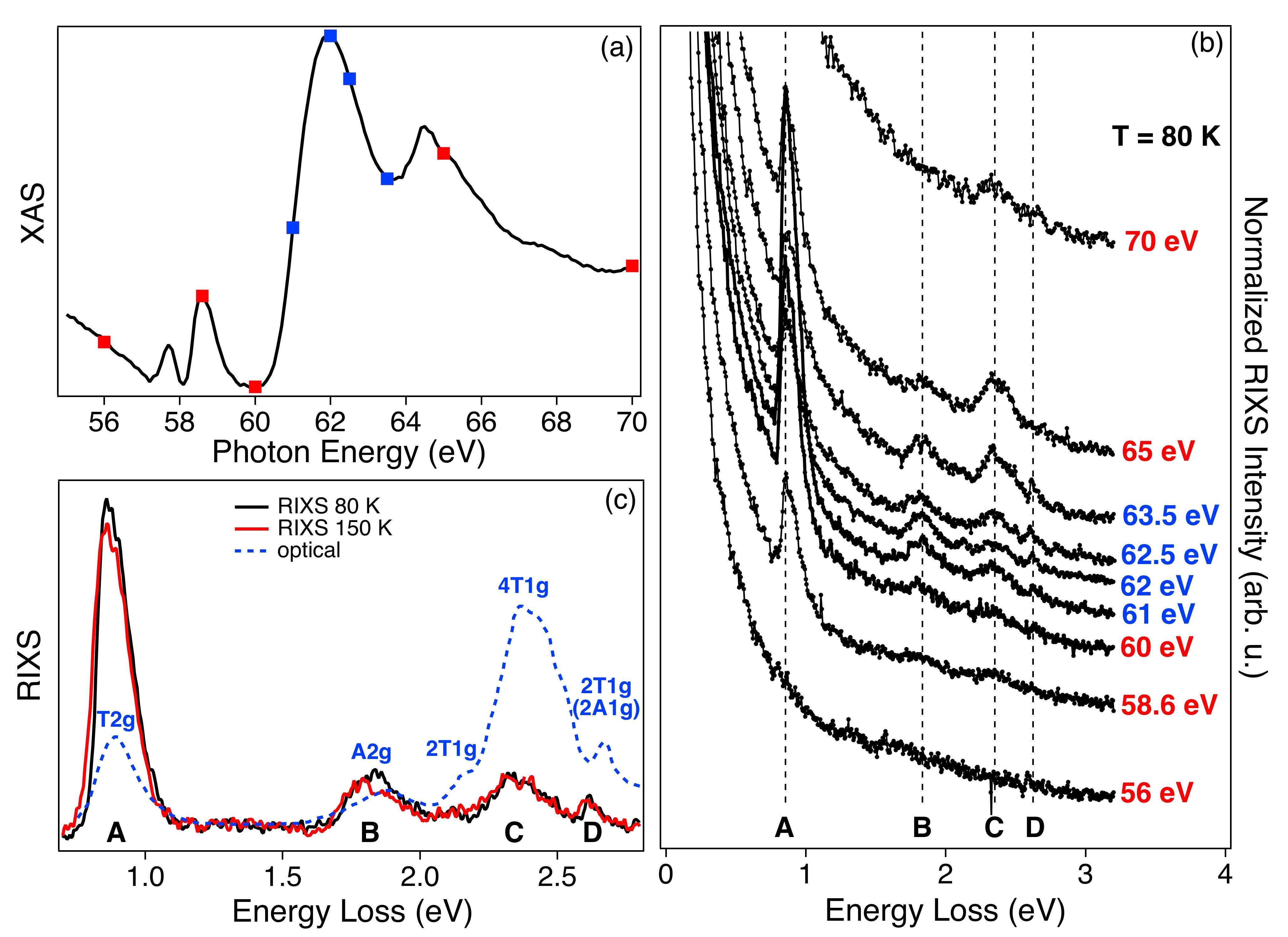}
\caption{(a) Co M$_{23}$ edge XAS spectrum recorded in total electron yield (TEY) mode at 80 K. (b) Stacked plot of Co M$_{23}$ edge RIXS spectra measured at 80 K. The RIXS spectra are plotted as a function of energy loss. The excitation photon energies are listed next to the curves and also denoted by the red and blue squares in panel (a). The strong elastic peak at zero energy loss is not fully displayed. (c) Comparison of two RIXS spectra measured above (150 K) and below (80 K) the N\'{e}el temperature. Each spectrum at respective temperature is produced by summing individual spectra taken at excitation photon energies marked by the blue squares in panel (a). The blue dashed line is the optical absorption spectrum of KCoF$_3$ reproduced from \citet{Ferguson:1963gz}. The main visible absorption bands are labelled in the figure.}
\label{fig2}
\end{figure}

\section{experiments}
Co M$_{23}$-edge X-ray absorption spectroscopy (XAS) and RIXS spectra \cc{of a pure KCoF$_3$ single crystal} were acquired at beamline 4.0.3 (MERLIN) RIXS endstation (MERIXS)\citep{Chuang:2012he} at the Advanced Light Source (ALS), Lawrence Berkeley National Laboratory (LBNL). RIXS data were recorded using the slitless VLS based X-ray emission spectrometer equipped with a commercially available in vacuum CCD detector\cite{Chuang:2012he,Chuang:2005ex}. \cc{The sample and the spectrometer were placed at 20$^{\circ}$ grazing incidence angle and 90$^{\circ}$ relative to the incoming X-ray beam, respectively. The photon polarization was maintained in the scattering plane in order to minimize the elastic scattered radiation. XAS measurements were recorded in the total electron yield (TEY).}
The atomic multiplet calculation was performed using Kramers-Heisenberg formalism\cite{Wray:2015htb,Wray:2012dca,Wray:2013goa,Wray:2012wz}. The Slater-Condon parameters were renormalized to 75\% of the Hartree-Fock values for 3d-3d interactions, and to 67\% for 3p-3d interactions. Spin-orbit couplings for core and valence levels were not renormalized. The crystal field strength was set to 10Dq=0.90 eV at T=0 K. The temperature effect was accounted for by applying a correction factor of $(1+T/20000K)^{-5/3}$. The standard -5/3 exponent\cite{weckhuysen2000spectroscopy} and a volumetric thermal expansion coefficient were estimated from \citet{Ratuszna:1979ct}. \cc{The ground states in the scattering equation were a Boltzmann weighted ensemble of low energy multiplet states. This weighting is the origin of the pseudo-anti-Stokes effect, in which RIXS features become broader on the energy gain-side at higher temperatures}. The intensity ratio between the elastic and inelastic features is around 10$^4$, similar to previous M-edge RIXS results on other compounds\citep{Wray:2013go,Kuiper:1998il,Chiuzbaian:2005ifa}. Because of that, background curves produced by polynomial fitting to the experimental RIXS spectra in [0.5 eV, 1.55 eV], [1.15 eV, 3.2 eV] energy windows were used for subtraction to reveal the weak inelastic features in Figs. \ref{fig2}(c) and \ref{fig3}(a-b). 
All inelastic features occurred at constant energy loss, and no charge transfer features were observed in the spectra. 
In order to retrieve the peak energy positions, the spectral features A, B, C and D were fitted with a set of Voigt functions.  
The Gaussian width of the Voigt is fixed at 15 meV, which was evaluated from the lowest temperature rising edge of peak A.

\section{results and discussion}
The X-ray absorption spectrum recorded in total electron yield mode around Co M$_{23}$-edge at 80 K is shown in Fig.\ref{fig2}(a).
The XAS spectrum exhibits a prominent double-peak structure from Co 3p$_{3/2,1/2}$ core levels in 61-67 eV energy range and two well-resolved pre-edge features in 57-59.5 eV energy range. The overall XAS lineshape bears similarity to that of CoO, suggesting the comparable energy scales for these two materials despite their different crystallographic structures. The experimental RIXS spectra at 80 K are shown in Fig. \ref{fig2}(b), with their excitation photon energies listed next to the curves and labeled by red and blue squares in Fig. \ref{fig2}(a). These spectra are dominated by the strong elastic peak at zero energy loss, which is not fully displayed in this figure; however, when zooming in the low intensity tail region around the elastic peak, four distinct inelastic features labeled A, B, C, and D can be identified in the [0.8 eV, 2.9 eV] energy loss window. These features show strong resonance enhancement when excitation photon energies are tuned to Co M$_3$ resonance (blue squares in Fig. \ref{fig2}(a)), confirming their origin from Co 3d orbitals. They also resemble the ones in a similar system CoO\citep{Wray:2013go}. But due to higher ionicity of KCoF$_3$ , features B and C are more pronounced in this material and we also observe a very sharp inelastic feature at higher energy loss (feature D, 27 meV FWHM) that is absent in the CoO spectra. 

\begin{figure}[t!]
\captionsetup{justification=centerlast}
\centering
\includegraphics[width=1\columnwidth]{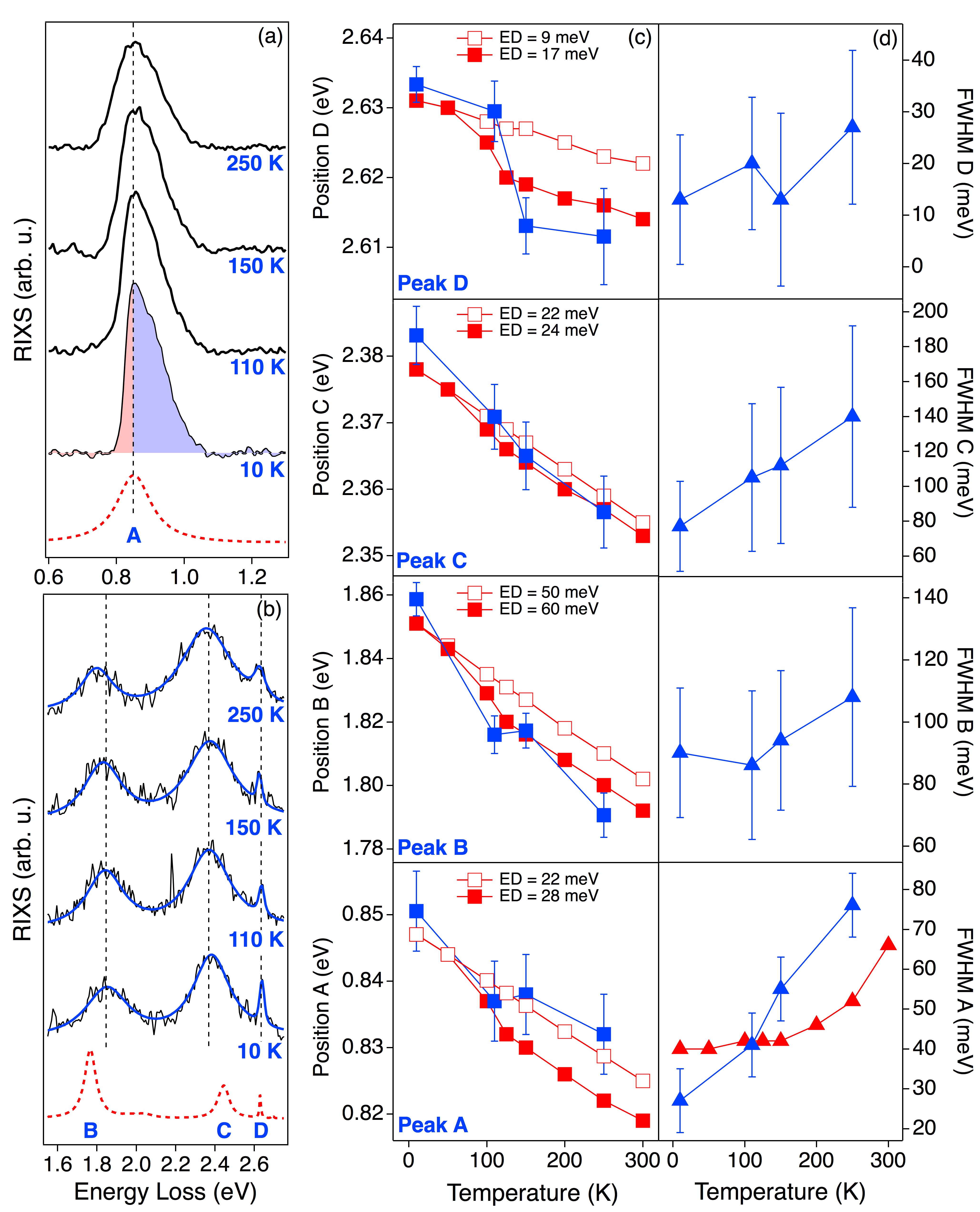}
\caption{(a, b) dd excitations (features A, B, C, and D) as a function of sample temperature. The RIXS spectra at selected temperatures (black curves) are plotted after subtracting the background (see text). The red dashed line is the theoretical RIXS spectrum. The high and low energy loss sides of feature A are marked by blue and red shades respectively in panel (a) (see text). The blue lines in panel (b) are the fitting functions overlaid on top of the RIXS spectra (see text for the details of fitting functions). (c) Temperature dependence of experimental (blue squares) and theoretical (red squares) peak positions. Experimental peak positions are determined from the fitting of experimental RIXS spectra. Two theoretical curves are shown in each sub-panel: the curve with red filled squares considers the mean field thermalization of near-neighbour magnetic moments, whereas the curve with red open squares fixes the near-neighbour magnetic moments to their T = 10 K value. Both curves include the effects of lattice expansion and thermalization of electronic states. Numbers in meV in the figure show the overall energy dispersion (ED) of each feature with respect to temperature. (d) Temperature dependence of experimental (blue triangles) and theoretical (red triangles; only for feature A) FWHM of the leading edge of four RIXS features. }
\label{fig3}
\end{figure}

In Fig. \ref{fig2}(c), we show two RIXS spectra measured at 150 K (red line) and 80 K (black line), above and below T$_N$. Each spectrum is obtained by summing individual spectra measured at excitation photon energies around Co M$_{3}$ edge (blue squares in Fig. \ref{fig2}(a)) at respective temperature. The four main RIXS features can be correlated with their counterparts in the optical absorption spectrum in visible and near infrared regimes\citep{Ferguson:1963gz} (blue dashed line): features A, B, C, and D correspond to excitations with T$_{2g}$, A$_{2g}$, $^{4}$T$_{1g}$, and $^{2}$T$_{1g}$ symmetries. The comparison between them shows that the high temperature spectrum exhibits a shallower slope on the energy gain side of all inelastic features, consistent with the “pseudo-anti-Stokes” effect identified in CoO\citep{Wray:2013go}.
This pseudo-anti-Stokes effect is particularly visible in feature A (see Fig. \ref{fig3}(a)) because in the atomic multiple calculation, the multiplets forming this feature are confined only within $\sim$20 meV energy range at 0 K. Thus as temperature is increased, the thermally excited states involved in RIXS process manifest additional spectral weight in the energy gain side (for example, the red shaded area in Fig. \ref{fig3}(a)) that increases with temperature. 
Besides this pseudo-anti-Stokes effect, one can see that the lineshape of feature A is highly asymmetric with an extended tail on the higher energy loss side, which gives the full width at half maximum (FWHM) around 120 meV at 10 K (for example, see the blue shaded area in Fig. \ref{fig3}(a)). This tail suggests the presence of other contribution from shake-up process. Unlike in the case of CoO where strong nearest-neighbor magnetic interaction along [111] direction was proposed to explain this energy loss tail, there are no nearest-neighbours along [111] direction in this perovskite compound. On the other hand, the high ionicity of KCoF$_{3}$ may enhance the contribution from lattice degree of freedom, as in the RIXS process where the occupancy of poorly screened Co 3d orbitals is changed, the Co-F bond length can be affected accordingly to lead to the phonon shake-up (see later discussion). 

In that regard, we follow the approach in Ref. \citep{Wray:2013go} to simulate the RIXS spectra with the consideration of pseudo-anti-Stokes effect on the energy gain side and the contribution from inter-atomic many-body dynamics on the energy loss side of excitations. A major contribution to the energy loss shake-up part of spectra is a continuum-like distribution of states spread over $\sim$150 meV, which likely represents simultaneous excitations of multi-phonons with matrix elements that shrink upon increasing the number of phonons. We also adopt the self-consistent mean-field approach for the temperature dependence of magnetic interaction. In the mean field theory, the 115.3 K N\'{e}el temperature implies an exchange field of J = 10.56 cm$^{-1}$ between neighboring Co sites, which is in agreement with the 10.3-10.6 cm$^{-1}$ values reported in Raman and neutron studies\citep{Moch:1973jd}\citep{Buyers:1971kxa}. The local exchange field is set equal to the mean field value from the six antiferromagnetically coupled neighbouring Co atoms. Representative simulated RIXS spectra are shown as red dashed lines in Figs. \ref{fig3}(a) and \ref{fig3}(b). 

Following the 3d$^7$ Tanabe-Sugano diagram and in agreement with the assignment in \citet{Ferguson:1963gz}, the best candidate for the 2.6 eV feature is a $^2$T$_{1g}$-symmetry feature, that comes from a spin $\frac{3}{2}$ to $\frac{1}{2}$ high spin to low spin transition.  There are actually two excitations very close by in energy ($^2$T and $^2$A symmetry), both of which are high spin to low spin excitations. However the $^2$A state shows negligible RIXS intensity in our calculation.

To quantify the temperature dependence of elementary excitations shown in Figs. \ref{fig3}(a) and \ref{fig3}(b), we carried out the fitting procedure outlined in the experimental sections to determine the peak positions for all the features. Figure \ref{fig3}(c) shows the experimental (blue squares) and theoretical (red open and filled squares; from atomic multiple calculation) temperature dependence of the peak positions. Two theoretical curves are shown in each sub-panel: the curve with red filled squares takes the mean field thermalization of nearest-neighbour magnetic moments into account, whereas the curve with red open squares has fixed nearest-neighbour magnetic moments at their T=10 K value (assuming the magnetic order is preserved over the whole temperature range). Both curves include the effects of lattice expansion\citep{Ratuszna:1979ct} and thermalization of electronic states.

The theoretical curves reproduce the observed temperature dependence of experimental peak positions nicely, implying that the lattice expansion and thermalization of electronic states are essential ingredients for simulating the RIXS spectra. However, there are subtle differences. As can be seen from the comparison of two theoretical curves (open and filled red squared curves in Fig. \ref{fig3}(c)), the presence of antiferromagnetic ordering affects the energy shift. However, the effect of magnetic ordering is different for different features. For example, for features B and C, excluding the antiferromagnetic ordering above T$_{N}$ only slightly increases dispersion (10 meV increase over 60 meV for feature B and 2 meV increase over 24 meV for feature C), and the effect for feature D is much larger (8 meV increase over 17 meV). Within the model, the relative large discontinuity of the energy position of feature D across the magnetic transition occurs because this is a high spin to low spin excitation (S=3/2 to S=1/2), which intrinsically weakens the local antiferromagnetic correlations.

Generally, the magnetic part of the total energy shifts is less than $\frac{1}{6}$ of the lattice part, upon heating from 10 to 150 K. Above 150 K, only the lattice part matters in a mean-field picture, though in reality a magnetic contribution due to short range exchange interaction continues to play a role. The major exception to this picture is the 2.6 eV feature, where the magnetism contributes almost 50\% of the total energy shift. Notably, the highly resolved experimental data clearly show a $\sim$20 meV discontinuity in the energy shift of that feature when crossing the magnetic transition temperature. This is corroborated by calculated value of $\sim$ 17 meV in our mean field many-body modelling of the magnetic interaction between neighbouring Co sites.

\begin{figure}[t!]
\captionsetup{justification=centerlast}
\centering
\includegraphics[width=1\columnwidth]{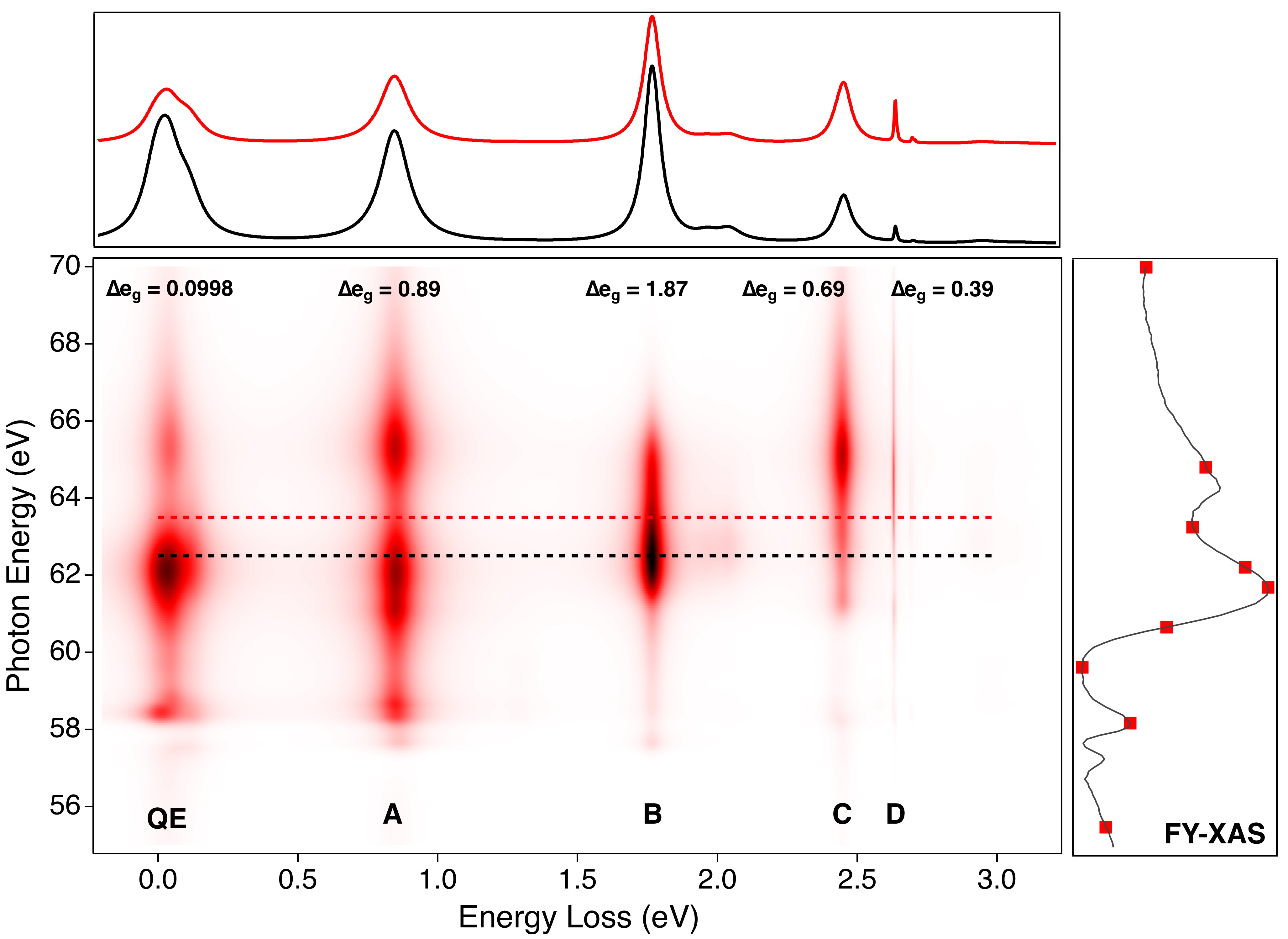}
\caption{The theoretical RIXS map at 10 K around Co M$_{23}$ edge, with 20 meV spectral broadening applied to account for the experimental energy resolution. Two theoretical RIXS spectra at 63.5 eV (red curve) and 62.5 eV (black curve) excitation photon energies are shown on top panel. The experimental XAS spectrum is shown on right panel. The charge density variation $\Delta$e$_g$ for each principal excitation are listed in the figure.}
\label{fig4}
\end{figure}

The full theoretical RIXS map across Co M$_{23}$ edge at low temperature, with broadening of spectral features to account for the experimental resolution, is shown in Fig. \ref{fig4}. In the same figure, we also list the amount of charge density that is transferred into the e$_g$ states, relative to the ground state, for each excitation (averaged over incident energy). The change of occupation numbers $\Delta$e$_g$ give a qualitative understanding of how strongly the excitation couples to the shake-up phonons. This can be understood from the fact that reducing the number of e$_g$ electrons will reduce the equilibrium distance between Co and the neighbouring F, as well as the capability to shield the core hole created in the RIXS process. If this distance is greatly changed upon creating excitations in the RIXS process, the quenching of core hole is expected to produce multi-phonons. It is important to note that this kind of phonon coupling only matters for phonons created in the RIXS process and does not depend strongly on temperature. Correspondingly, $\Delta$e$_g$ only have a connection to the features width FWHM at low temperature. \cc{This effect is discussed for the transition metal L-edge in  \citet{Lee:2014ib}.}

Finally, the thermal dispersion of the inelastic features increases monotonically with the amount of $\Delta$e$_{g}$ electron density, because e$_{g}$ electron orbitals directly overlap with the fluorine sigma orbitals.
This picture can explain the remarkable energy stability of the lattice component of the 2.6 eV feature energy, (i.e. disregarding the discontinuity across the magnetic transition).

In conclusion, we report a comprehensive high resolved EUV-RIXS investigation of the excitation spectrum of orbital and spin degrees of freedom of antiferromagnet KCoF$_3$ revealing a binding energy shift of these excitations across the antiferromagnetic phase transition. 
Interpretation from the perspective of short-range magnetic energetics suggests that this energy shift arises through the cobalt high spin to low spin flips, which weakens the local antiferromagnetic correlations. 
It follows that, owing to the unquenched spin-orbit interaction of cobalt, the energy stability of the spin and orbital states is affected by the local spin excitations when the antiferromagnetic order is established. 
More in general, our results unveils the effect of the magnetic exchange energetics on the low energy landscape in this class of materials.

This work was supported by EU within FP7 project PASTRY [GA 317764]. The FERMI project at Elettra Sincrotrone Trieste is supported by MIUR under Grants No. FIRB RBAP045JF2 and No. FIRB RBAP06AWK3. The Advanced Light Source is supported by the Director, Office of Science, Office of Basic Energy Sciences, of the U.S. Department of Energy under Contract No. DE-AC02-05CH11231. Support by the Deutsche Forschungsgemeinschaft through the collaborative research center SFB925/project B2 is gratefully acknowledged.

\newpage

%

\end{document}